# Mechanisms of pyroelectricity in three- and two-dimensional materials

Jian Liu[1,*] and Sokrates T. Pantelides[1,2]

[1]Department of Physics and Astronomy, Vanderbilt University, Nashville, TN 37235, USA

[2]Department of Electrical Engineering and Computer Science, Vanderbilt University, Nashville, TN 37235, USA

Pyroelectricity is a very promising phenomenon in three- and two-dimensional (2D) materials, but first-principles calculations have not so far been used to elucidate the underlying mechanisms. Here we report density-functional theory (DFT) calculations based on the Born-Szigeti theory of pyroelectricity, by combining fundamental thermodynamics and the modern theory of polarization. We find satisfactory agreement with experimental data in the case of bulk benchmark materials, showing that the so-called electron-phonon renormalization, whose contribution has been viewed as negligible, is important. We predict out-of-plane pyroelectricity in the recently synthesized Janus MoSSe monolayer and in-plane pyroelectricity in the group-IV monochalcogenide GeS monolayer. It is notable that the so-called secondary pyroelectricity is found to be dominant in GeS monolayer. The present work opens a theoretical route to study the pyroelectric effect using DFT and provides a valuable tool in the search for new candidates for pyroelectric applications.

**Corresponding Author**
[*]Email: Jian.Liu@alumni.stonybrook.edu.

The introduction and implementation of the modern theory of polarization[1-3] have revolutionized first-principles studies of ferroelectric and piezoelectric materials. The pyroelectric effect[4-6] is the response of the spontaneous polarization $P_S$ with respect to temperature variations. It impacts a wide range of applications, including temperature sensors[7,8], nanogenerators[9,10], pyro-based energy harvesting[11], neutron generators[12], and self-powered ultraviolet nanosensors[13]. Non-ferroelectric pyroelectrics (e.g. wurtzite zinc oxide ZnO) are emerging as promising alternative candidates for pyroelectric devices[9,13] as they do not have a Curie temperature above which the pyroelectric behavior vanishes. The converse of the pyroelectric effect, the electrocaloric effect, has also attracted intensive interest[14-20]. Despite these extensive applications, the theory of the pyroelectric effect has not received due attention and, as a result, adequate understanding of the microscopic mechanisms that control the effect is still lacking.

A theory of pyroelectricity was constructed by Born[21] and Szigeti[22]. Two main contributions were identified: The primary pyroelectricity (at constant external strain) and the secondary pyroelectricity (associated with thermal expansion). The secondary pyroelectricity is generally much smaller than the primary pyroelectricity[4]. One often-mentioned exception is the case of tourmaline in which the secondary pyroelectricity dominates[4]. Furthermore, the primary contribution was divided into two parts[21,22]: contributions from the mean displacements of atoms as rigid ions and from the electronic redistribution caused by thermal vibrations (known as electron-phonon renormalization). The former, i.e. the rigid-ion model in which an effective charge is assigned to each ion rigidly, has been widely adopted in model-potential simulations[23-25] and analysis of experimental data[26,27] since it allows a simple interpretation of the pyroelectric behavior. So far, only one paper[28] has reported first-principles calculations of the rigid-ion contribution to the primary pyroelectricity based on the Born-Szigeti theory.

Very recently, theoretical predictions and experimental observations of two-dimensional (2D) ferroelectricity[29-33] and piezoelectricity[34-38] have emerged. As all ferroelectrics are pyroelectric (while all pyroelectrics are piezoelectric)[39], pyroelectricity is expected in 2D materials. Among various 2D materials, monochalcogenides have emerged as an important family for their large spontaneous polarization[29], ferroelectricity and piezoelectricity[40-42], ferroelasticity[43], second harmonic generation[44], photostriction[45] and bulk photovoltaic effect[46]. Meanwhile, the recently synthesized Janus MoSSe monolayer[47] has been predicted to exhibit out-of-plane piezoelectricity[48], which is highly desired for circuit designs, as was demonstrated for layered $\alpha$-$In_2Se_3$ nanoflakes[49,50]. First-principles calculations of pyroelectricity, however, have been very limited. Car-Parrinello molecular dynamics (CPMD) simulations have been reported[40,51], focusing mostly on the order-disorder transition in 2D ferroelectric materials, which, among other phenomena, induces pyroelectricity.

In this paper, by combining fundamental thermodynamics and the modern theory of polarization, we report comprehensive density-functional theory (DFT) calculations of the pyroelectric effect based on the Born-Szigeti theory of pyroelectricity. Wurtzite-structure ZnO and GaN are studied as benchmarks. We find satisfactory agreement with the experimental data, and show that substantial contributions to the primary pyroelectricity arise from electron-phonon renormalization, which has generally been viewed as negligible. For 2D materials, we demonstrate out-of-plane pyroelectricity in the recently synthesized Janus MoSSe monolayer and in-plane pyroelectricity in the group-IV monochalcogenide GeS monolayer. The former is particularly suitable for device applications, while the latter exhibits effective pyroelectricity that is orders of magnitudes larger than bulk pyroelectrics. Unexpectedly, the secondary pyroelectricity is found to be significant in monolayer GeS, contrary to the general belief for bulk prototypical pyroelectrics. The present work provides a theoretical route to study the pyroelectric effect that facilitates the search for 2D materials with potential pyroelectric applications.

According to the Born-Szigeti theory of pyroelectricity, the total pyroelectricity at constant stress σ is decomposed into two parts[4,21,22]: the primary (at constant external strain ε, labeled as $p_1$) and the secondary (associated with thermal expansion, labeled as $p_2$)

$$\left(\frac{dP_S}{dT}\right)_\sigma = p_1 + p_2 = \left(\frac{\partial P_S}{\partial T}\right)_\varepsilon + \sum_i \left(\frac{\partial P_S}{\partial \varepsilon_i}\right)_T \left(\frac{\partial \varepsilon_i}{\partial T}\right)_\sigma. \tag{1}$$

The secondary part corresponds to thermal-expansion-induced pyroelectricity, which can be readily obtained from the piezoelectric stress constants and the thermal expansion coefficients[4]. The primary part accounts for the "clamped-lattice" pyroelectricity when the external lattice parameters are held fixed. Expanding $P_S(T)$ in terms of normal mode amplitudes $Q$[21,22], one obtains

$$p_1 = p_1^{(1)} + p_1^{(2)} = \sum_j \frac{\partial P_S}{\partial Q_{\vec{0}j}} \frac{d\langle Q_{\vec{0}j}\rangle}{dT} + \sum_{\vec{q}\lambda} \frac{\partial^2 P_S}{\partial Q_{\vec{q}\lambda}^2} \frac{d\langle Q_{\vec{q}\lambda}^2\rangle}{dT}, \tag{2}$$

where the brackets denote a thermal average. In the first-order primary pyroelectricity (labeled as $p_1^{(1)}$, which we identified earlier as the rigid-ion contribution), the summation runs over $\vec{q} = 0$ optical modes of symmetries that allow a non-vanishing first-order static effect ($\langle Q_{\vec{0}j}\rangle \neq 0$). $\langle Q_{\vec{0}j}\rangle$ can be calculated using the quasi-harmonic approximation (QHA)[28],

$$\langle Q_{\vec{0}j}\rangle = -\sum_{\vec{q}\lambda} \frac{\hbar}{2} \frac{2n_{\vec{q}\lambda}+1}{\omega_{\vec{0}j}^2} \frac{\partial \omega_{\vec{q}\lambda}}{\partial Q_{\vec{0}j}}, \tag{3}$$

where ω is the phonon eigenfrequency and n is the Bose-Einstein distribution function. In the second-order primary pyroelectricity (labeled as $p_1^{(2)}$, which we identified earlier as the electron-phonon renormalization contribution), all phonon modes contribute by their mean-square displacement

$$\langle Q_{\vec{q}\lambda}^2\rangle = \frac{\hbar}{2} \frac{2n_{\vec{q}\lambda}+1}{\omega_{\vec{q}\lambda}}. \tag{4}$$

It is then understood that $p_1^{(1)}$ represents the rigid-ion pyroelectricity induced by the mean displacement of ions carrying Born effective charges $Z^*$, while $p_1^{(2)}$ describes the electronic redistribution as atoms vibrate around the equilibrium.

All ingredients in the Born-Szigeti theory of pyroelectricity are readily accessible from the elementary theories of thermodynamics and spontaneous polarization. Details of the computational methods are given in Supplementary Materials.

Wurtzite structure is the highest-symmetry structure that exhibits spontaneous polarization[52], with three structural parameters a, c, and u. Computed properties of ZnO are given in Table 1. The calculated structural properties reproduce experimental values well, within the usual DFT error. As already reported, the piezoelectric constants are delicate quantities and their calculations are sensitive to many numerical parameters[53]. ZnO has a high melting temperature of 2,248 K. We calculate the pyroelectric coefficients in the temperature range well below the melting temperature, where QHA applies. Above 1000 K, one may expect deviations due to anharmonic effects. Due to the zero-point effect, the structural parameters renormalize at zero temperature. These are (0.0026, 0.0019, 0.00025) for ($\Delta a/a_0$, $\Delta c/c_0$, $\Delta u/u_0$). Polarization renormalizes accordingly. For ZnO, the zero-point renormalization (labeled as $\Delta P_0$) is -0.0015 $C/m^2$, negligibly small compared to $P_s$. The computed pyroelectric coefficients for ZnO and GaN are shown in Fig. 1. Compared with experimental data, excellent agreement is found. Compared with the first-order term $p_1^{(1)}$, the second-order term $p_1^{(2)}$ makes significant and equally important contributions, which is somewhat unexpected.

Table 1. Calculated internal (dimensionless) and external (in units of Å) structural parameters, spontaneous polarization and its zero-point motion correction (in units of $C/m^2$), piezoelectric constants (in units of $C/m^2$), Born effective charge (in units of e), and thermal expansion coefficients at 300 K (in units of $10^{-6}K^{-1}$), for ZnO.

| ZnO | a | c | u | $\alpha_a$ | $\alpha_c$ |
|---|---|---|---|---|---|
| present | 3.20 | 5.17 | 0.379 | 4.4 | 3.0 |
| previous/expt. | 3.247[a] | 5.204[a] | 0.382[b] | 4.31[a] | 2.49[a] |
| | $P_s$ | $\Delta P_0$ | $Z^*$ | $e_{31}$ | $e_{33}$ |
| present | -0.035 | -0.0015 | -2.17 | -0.57 | 1.32 |
| previous/expt. | -0.029[c], -0.057[d] | | -2.08[c], -2.11[d], -2.1[e] | -0.69[c], -0.51[d] | 1.31[c], 0.89[d] |

[a] Ref. [54] [b] Ref. [26] [c] Ref. [53] [d] Ref. [55] [e] Ref. [56]

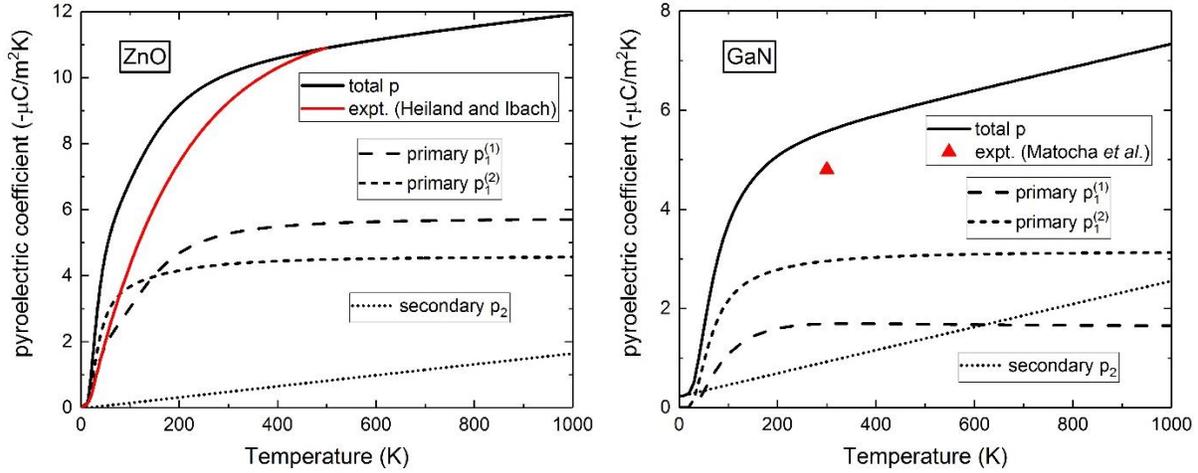

Figure 1. Calculated pyroelectric coefficients of ZnO and GaN. Experimental data for ZnO and GaN are from Ref. [57] and Ref. [58] respectively.

Starting with the first-order primary pyroelectricity, the relevant $A_1$ phonon mode has opposite displacements of cations and anions along the polar direction, which gives the internal structural parameter u. For ZnO, experimental data on the temperature dependence of u are scarce and contradictory[26,59]. Albertsson and Abrahams[26] found that $u(T)$ remains unchanged between 20 and 300 K, while above 300 K their measurements exhibit a much larger internal thermal expansion compared to the present calculations. We are not able to identify the source of the discrepancy. It is possible to construct an effective one-particle potential (OPP) derived up to cubic anharmonicity for wurtzite crystals[60] (see Supplementary Materials). We interpolate the potential parameters with a $3 \times 3 \times 2$ (72-atom) supercell. The effective OPP reproduces well the trend and the magnitude of $u(T)$ calculated with the application of QHA, as shown in the inset of Fig. 2a. This validates the adequacy of Eq. (3) in treating mean atomic thermal displacement (internal thermal expansion).

For the second-order primary pyroelectricity, the symmetry-preserving $E_2^{(low)}$ mode exhibits no mean atomic thermal displacement ($\langle Q \rangle = 0$), and therefore $p_1^{(1)}$ vanishes for this mode. However, the thermal vibration of this mode induces deformation of the electronic cloud, giving rise to $p_1^{(2)}$, as shown in Fig. 2b. Its second-order nature (with respect to normal mode amplitude) is clearly manifested in the parabolic shape, consistent with the fact that for ZnO the off-diagonal components of the Born effective charge tensor are zero. It is worth mentioning that Born suggested (without a better alternative) that the second-order polarization expansion coefficients $\partial^2 P_S / \partial Q_{\vec{q}\lambda}^2$ could be replaced by some constant average[21]. The present calculations find that these coefficients largely vary with wavevector $\vec{q}$ and mode $\lambda$.

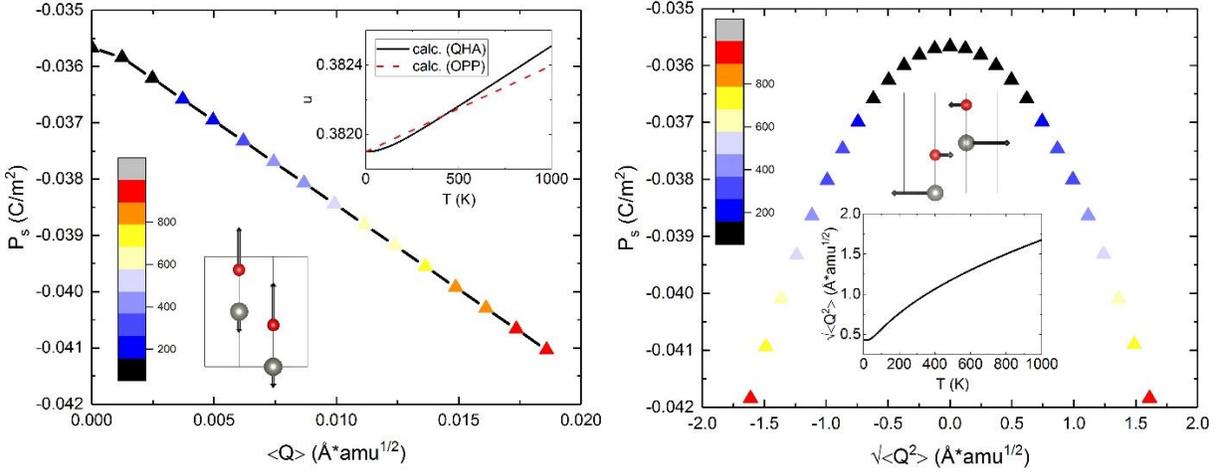

Figure 2. Primary pyroelectric effect in ZnO. Polarizations are evaluated with atoms statically displaced along the normal mode Q over a range of "frozen-in" amplitudes. Arrows attached on each atom indicate the eigenvectors of the phonon mode. Left panel: first-order $p_1^{(1)}$. Color indicates the temperature corresponding to the mean atomic thermal displacement. Th einset shows the calculated internal thermal expansion $u(T)$, compared with results obtained by one-particle potential (OPP). Right panel: second-order $p_1^{(2)}$. Color indicates the temperature corresponding to the amplitude of thermal vibrations. The inset shows the mean squared displacement of the $E_2^{(low)}$ phonon mode.

Moving to the secondary pyroelectricity, thermal expansion of the structural parameters $a(T)$ and $c(T)$ of ZnO have been measured for a wide range of temperatures[54,61,62]. ZnO exhibits negative thermal expansion (NTE) at low T. Ibach[62] found that the thermal expansion coefficients of ZnO remain negative below 80 K and 120 K for a and c respectively. Reeber[61] found that a is minimum at 93 K, while a definite temperature for the NTE of c cannot be assigned due to the limited precision of the experiment. Overall, the present calculations yield satisfactory agreement with the experimental data. NTE is predicted for a and c below 130 K. Note, however, that the present calculations apply the "zero static internal stress" approximation (ZSISA)[63], while the full application of QHA requires treating both external and internal strains on the same footing[64]. The uncertainties of ZSISA will be discussed in a future work.

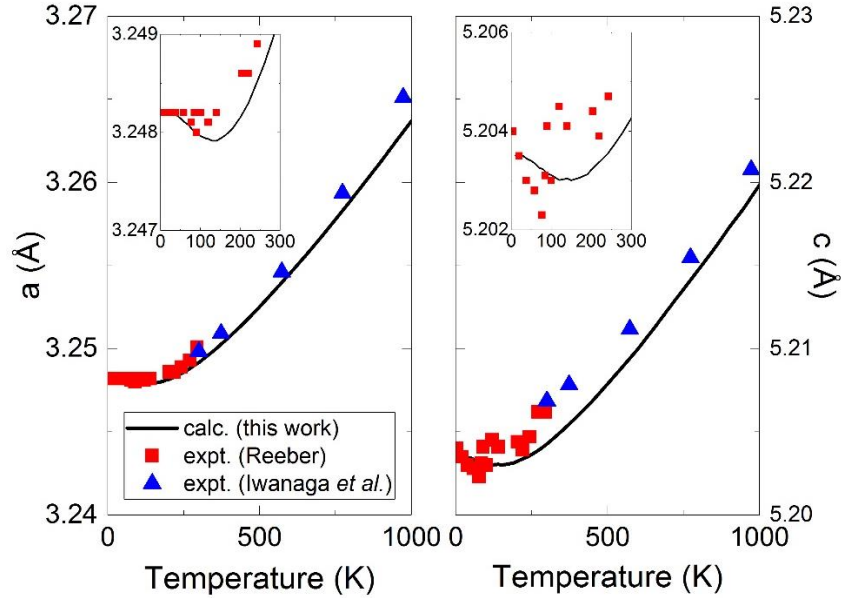

Figure 3. Calculated thermal expansion of ZnO with the application of ZSISA. Experimental data for a and c are from Ref. [61] (4.2-300K) and Ref. [54] (300-900 K) respectively. To stress the agreement in the temperature dependence, the calculated lattice constants at 0 K (including zero-point motion) are aligned to those measured at low temperatures.

Having demonstrated that pyroelectricity can be predicted accurately, we now extend our discussion to 2D monolayers. Below we demonstrate out-of-plane pyroelectricity in the recently synthesized Janus MoSSe monolayer and in-plane pyroelectricity in the group-IV monochalcogenide GeS monolayer. This material has been predicted to be stable, but has not yet been synthesized. The two materials are representatives of broad classes. Even though different pyroelectric behavior may be realized in other classes of 2D materials, the present calculations and analysis serve as prototypes for further studies. The Janus MoSSe monolayer[47,48] adopts a structure similar to $2H-MoS_2$, with each Mo atom being sixfold coordinated. The Mo-plane is sandwiched by the S- and Se- planes. As all pyroelectrics are piezoelectric, the Janus MoSSe monolayer allowing piezoelectricity in the out-of-plane direction[48] is also a potential candidate exhibiting out-of-plane pyroelectricity. The group-IV monochalcogenide GeS monolayer[29,51] adopts an orthorhombic structure similar to phosphorene with all atoms being threefold coordinated. For bulk monochalcogenides, the presence of inversion symmetry forbids pyroelectricity. In the monolayer, spontaneous polarization is allowed in the armchair direction due to symmetry breaking.

Computed pyroelectric coefficients of 2D monolayers are shown in Fig. 4. To facilitate a direct comparison, an effective monolayer thickness of 6.2 Å (5.2 Å) is assumed for MoSSe (GeS). The out-of-plane pyroelectricity exhibited in MoSSe monolayer is two orders of magnitude smaller than bulk ZnO, which is also manifested in the comparison of their piezoelectric coefficients ($|e_{31}| = 0.005$ and $0.5$ C/m$^2$ for MoSSe monolayer[48] and bulk ZnO[55] respectively). On the other hand, GeS monolayer exhibits large in-plane pyroelectricity (one order of magnitude larger than bulk ZnO). It is noteworthy that, the large pyroelectricity arises mainly from the secondary effect, which is contrary to the general expectation that primary pyroelectricity dominates over secondary pyroelectricity[4] (see Fig. 1a, where for ZnO the secondary pyroelectricity is negligibly small). Considering their comparable piezoelectric coefficients ($|e_{11}| = 0.88$ and $|e_{33}| = 0.89$ C/m$^2$ for GeS monolayer[41] and bulk ZnO[55] respectively), this contrast is totally attributed to the difference in their thermal expansion behaviors. Indeed, at room temperature, the calculated thermal expansion coefficients (in unit of $10^{-6}K^{-1}$) read $\alpha_a = -110$ and $\alpha_b = 64$ for GeS monolayer, significantly larger than those of bulk ZnO[54] ($\alpha_a = -4.3$ and $\alpha_c = 2.5$).

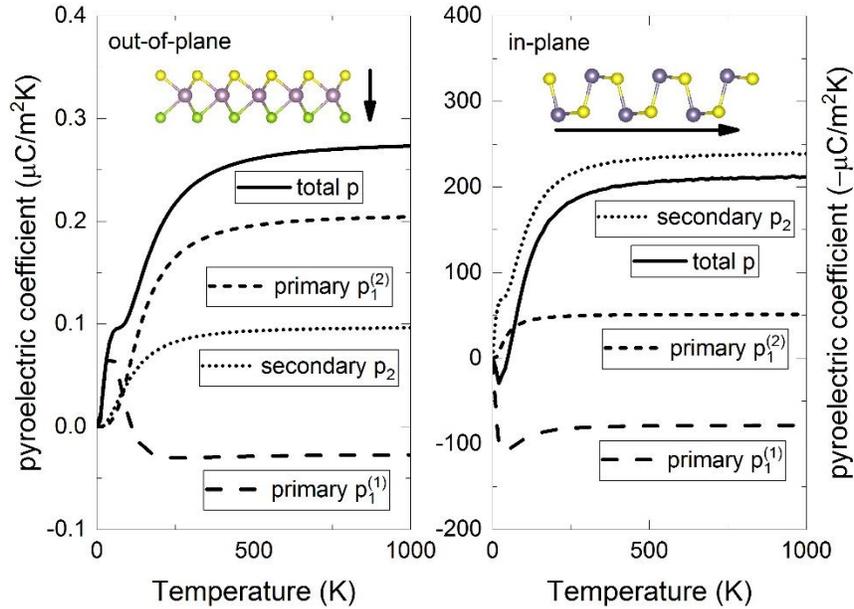

Figure 4. Calculated pyroelectric coefficients of MoSSe and GeS monolayers. Piezoelectric coefficients for MoSSe and GeS monolayers are from Ref. [48] and Ref. [41] respectively.

Alternative to the Born-Szigeti theory of pyroelectricity, the phenomenological Landau formalism is widely applied when the pyroelectric behavior near the phase transition is concerned. As done by Fei et al. in Ref. [29], for group-IV monochalcogenide monolayers, the potential energy is expanded in terms of polarization at each cell:

$$E = \sum_i \left(\frac{A}{2}P_i^2 + \frac{B}{4}P_i^4 + \frac{C}{6}P_i^6\right) + \frac{D}{2}\sum_{\langle i,j\rangle}(P_i - P_j)^2. \qquad (5)$$

The validity of this theory in various limits has been examined in detail in Ref. [65]. In the above formula, secondary (and electron-phonon renormalization as well) pyroelectricity is discarded, as the parameters A-D are fit to DFT calculations at T = 0 K. However, it has been shown that applying strain can dramatically tune the phase diagram of SnSe monolayers[29]. Furthermore, GeSe and SnSe monochalcogenide monolayers undergo a structural phase transition in which the rectangular lattice becomes square[40], which implies the importance of the coupling between the external elastic strains ($\varepsilon_a, \varepsilon_b$) and the internal local modes $P_i$. Here, with the only purpose of investigating the secondary pyroelectricity in the Landau formalism, we simply adopt this formula and allow the parameters A-D to depend on the lattice constants, as determined by free thermal expansion (calculated using the Grüneisen theory at the quasi-harmonic level[66-68]).

The calculated free thermal expansion is shown in Fig. 5b. With increasing temperature, the long axis a (in the armchair direction) lengthens continuously while the short axis b (in the zigzag direction) shortens continuously. A "structural phase transition" occurs when a and b become equal at 915 K, higher than that obtained from Car-Parrinello MD simulations[51] (510 K), while a similar trend (a drastic decrease of a/b) was found for the thermal expansion behavior[40,51]. Note that the lattice-constant curves in Fig. 5b continue past the crossing point due to the neglect of disorder in the Grüneisen theory and explicit anharmonic effects at the quasi-harmonic level. Once the cross, however, a phase transition occurs and the two lattice constants retain the same, as demonstrated in CPMD simulations[40,51]. The crossing of the two curves in Fig. 5b is analogous to $\Delta\alpha$ going to zero in Fig. 3c of Ref. [69].

As the lattice becomes increasingly strained, the double-well potential becomes increasingly shallower and narrower, giving rise to reduced ground-state spontaneous polarization and $T_c$ respectively, as shown in Fig. 5c-5d. Note that the deep blue line in Fig. 5d corresponds to the pyroelectricity (in Landau formalism) with lattice constants fixed at their zero-temperature values. Under free thermal expansion, the total pyroelectricity (shown by the black short-dot line in Fig. 5d) is enormously enhanced. Very recently, Barraza-Lopez et al. reached a similar conclusion (constraining the lattice constants to their magnitudes at zero temperature raises the transition temperature) through quantum MD simulations[69]. Considering the structural similarities, it is reasonable to expect this unusually large secondary pyroelectric response to be a general feature for the entire family of group-IV monochalcogenide monolayers.

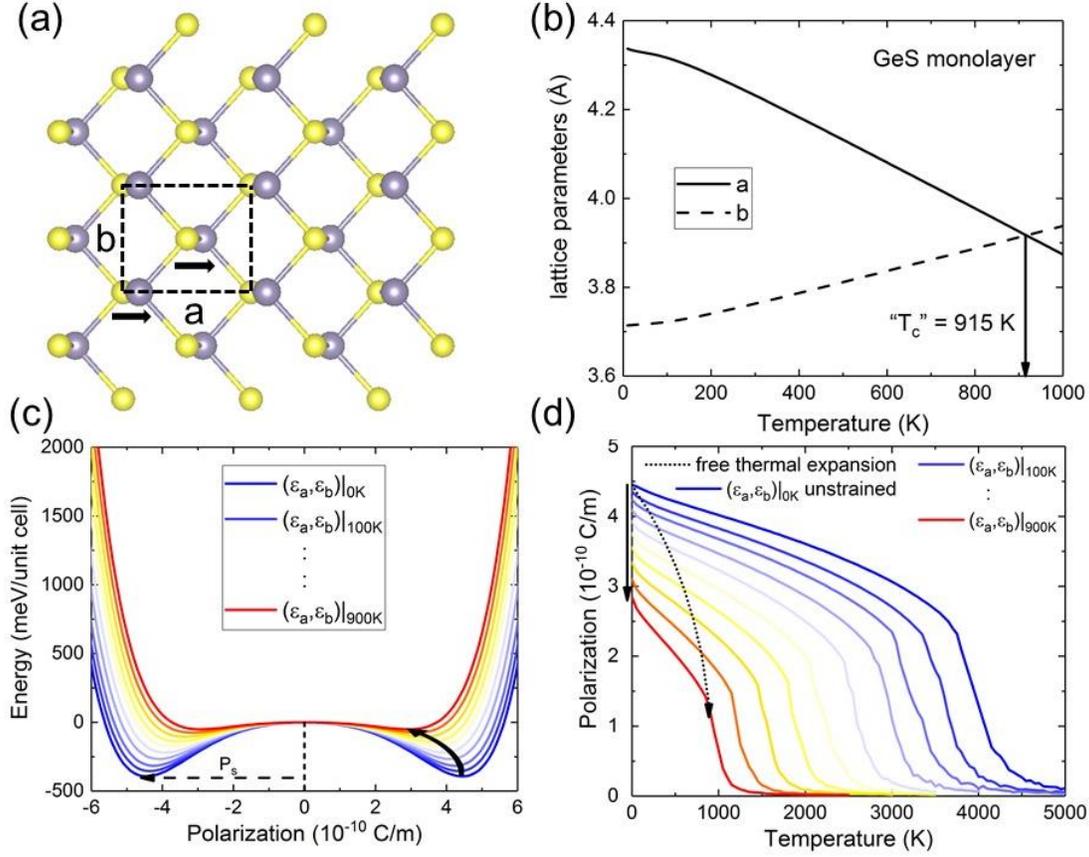

Figure 5. GeS monolayer: (a) Top view. The arrow shows the polarization along the armchair direction. (b) Thermal expansion calculated using the Grüneisen theory at the quasi-harmonic level. (c) Double-well potentials under thermal strains. The curved arrow indicates with increasing temperatures the reduction of both ground-state spontaneous polarization and potential barrier. (d) Pyroelectricity under thermal strains. The total pyroelectricity is denoted by the black short-dot line.

In the present study, we have primarily focused on "displacive" pyroelectricity (as in the Born-Szigeti theory). In the case of ferroelectric pyroelectrics such as the GeS monolayer, quantum MD simulations have demonstrated that, at finite temperature, the polarization in each unit cell (shown by black arrow in Fig. 5a) disorders in both magnitude and direction due to structural degeneracies[40,51]. Such disordered polarization gives rise to "order-disorder pyroelectricity". Note that this is to be distinguished from the pyroelectricity of the "displacive" type as in the Born-Szigeti theory.

In conclusion, following the Born-Szigeti theory of pyroelectricity, and by combining fundamental thermodynamics and the modern theory of polarization, we introduce a first-principles route allowing the computation of the pyroelectric response. A case study of bulk wurtzite-structure ZnO and GaN shows good agreement with experimental data. We demonstrate that substantial contributions to the primary pyroelectricity arise from electron-phonon renormalization. Applying this route to 2D materials, we demonstrate out-of-plane pyroelectricity in MoSSe monolayer and in-plane pyroelectricity in GeS monolayer. A large pyroelectric response is found in GeS monolayer, and is attributed to the secondary effect. Applying strains can dramatically tune the intrinsic pyroelectricity, while allowing free thermal expansion results in an enormous enhancement. These theoretical results may motivate future research interest into pyroelectricity with applications in energy harvesting and sensors.

## Acknowledgements

This work was supported by DOE grant DE-FG02-09ER46554 and the McMinn Endowment at Vanderbilt University. Computations were carried out at the National Energy Research Scientific Computing Center (NERSC), a DOE Office of Science User Facility supported by the Office of Science of the U.S. Department of Energy under Contract No. DE-AC02-05CH11231, and the Extreme Science and Engineering Discovery Environment (XSEDE), which is supported by National Science Foundation grant number ACI-1053575. This research also used computational resources of the Texas Advanced Computing Center (TACC) through XSEDE under project TG-DMR170102.